\documentstyle[11pt,aasms4]{article}

\lefthead{Worrall \& Birkinshaw}
\righthead{X-ray atmospheres of B2 radio galaxies}

\begin{document}

\title{X-ray-emitting Atmospheres of B2 Radio Galaxies}

\author{D.M.~Worrall \& M. Birkinshaw}
\affil{Department of Physics, University of Bristol, U.K.}
\affil{Harvard-Smithsonian Center for Astrophysics, Cambridge, MA 02138}

\authoremail{d.worrall@bristol.ac.uk}

\begin{abstract}

We report ROSAT PSPC spatial and spectral analysis of the eight B2
radio galaxies NGC~315, NGC~326, 4C~35.03, B2~0326+39, NGC~2484,
B2~1040+31, B2~1855+37, and 3C~449, expected to be representative of
the class of low-power radio galaxies.  Multiple X-ray components are
present in each, and the gas components have a wide range of linear
sizes and follow an extrapolation of the cluster X-ray
luminosity/temperature correlation, implying that there is no
relationship between the presence of a radio galaxy and the gas
fraction of the environment.  No large-scale cooling flows are found.
There is no correlation of radio-galaxy size with the scale or density
of the X-ray atmosphere.  This suggests that it is processes on scales
less than those of the overall gaseous environments which are the
major influence on radio-source dynamics.  The intergalactic medium is
usually sufficient to confine the outer parts of the radio structures,
in some cases even to within 5~kpc of the core.  In the case of
NGC~315, an extrapolation suggests that the pressure of the atmosphere
may match the minimum pressure in the radio source over a factor of
$\sim 40$ in linear size (a factor $\sim 1600$ in pressure).

\end{abstract}


\section{Introduction} \label{introsec}

Sensitive ROSAT observations have shown that low-power radio galaxies
display a mixture of X-ray emission components from a combination of
processes.  The dominant components are usually (a) an extended
atmosphere of X-ray emitting gas and (b) unresolved emission whose
primary origin is most likely to be non-thermal radiation associated
with the radio-emitting plasma in the innermost parts of the source
(\markcite{worr94}Worrall \& Birkinshaw 1994;
\markcite{hard99}Hardcastle \& Worrall 1999).  The balance of these
and other emission components differs from galaxy to galaxy, and each
source must be observed in detail in order to disentangle the emission
processes and probe the physics.

The X-rays associated with the most compact active regions of radio
galaxies are likely to originate not only as emission from the inner
jets, which are affected by relativistic beaming, but also as more
isotropically-emitted radiation produced even closer to the central
engine of the AGN.  Since radio galaxies do not shine as brightly in
compact X-ray emission as Seyfert galaxies and quasars, this AGN
component is suppressed, due either to low radiative efficiency of
accretion-related thermal emission or to asymmetric obscuration.  In
the latter case the AGN's contribution to the compact emission should
increase towards higher X-ray energies.  If models of source
unification are to be addressed in the X-ray, the extended non-nuclear
emission must be measured and removed to reveal the uncontaminated
nuclear properties of the active galaxy.

The existence of an extended X-ray emitting medium is probably
essential for prominent radio jets to develop, but to find trends in
the morphology of radio structures with gas density and distribution
requires studies of the atmospheres on a source-by-source basis.
Excluding roughly the inner kpc, it seems in general that the external
medium supplies more than enough pressure to confine the radio jets of
low-power radio galaxies, assuming a jet pressure calculated using
minimum-energy arguments (e.g., \markcite{morg88}Morganti et al.~1988;
\markcite{kill88}Killeen, Bicknell \& Ekers~1988;
\markcite{fer95}Feretti et al.~1995; \markcite{tru97}Trussoni et
al.~1997).  In some cases an apparent evacuation of the external
medium within the jets argues that additional jet pressure is required
and must be supplied by something other than thermal material (e.g.,
\markcite{boh93}B\"ohringer et al.~1993; \markcite{hard98}Hardcastle,
Worrall \& Birkinshaw~1998).  Such `light' jets should then be
affected by buoyancy forces (e.g., \markcite{wbc95}Worrall, Birkinshaw
\& Cameron~1995), and it is of interest to examine how morphological
features, such as the disruption of jets into lobe emission, map onto
abrupt or gradual changes in the X-ray emitting environment.  In
exceptional cases it appears that relatively low-powered jets on
scales of tens to hundreds of kpc remain overpressured with respect to
the external medium (e.g., \markcite{birk93}Birkinshaw \& Worrall
1993).

The X-ray emitting medium may play an important role in the fueling of
the central engine, particularly if it is involved in a cooling flow
that continues into the nuclear regions.  However, observations with
excellent spatial resolution are needed to distinguish the central
part of a cooling flow from the smaller-scale emission associated with
the inner radio jets or the AGN itself.

This paper presents a thorough analysis of eight low-power radio
galaxies using the best available X-ray data.  In \S\ref{samplesec} we
describe the selection of sources, the extent to which they are
representative of their parent B2-galaxy sample, and the ROSAT
observations. \S\ref{spatsec} and \S\ref{specsec} describe the spatial
and spectral distributions of the X-ray emission.  In \S\ref{physsec}
we discuss the physical characteristics of the X-ray emitting
environments and how their pressures compare with those of the
kpc-scale radio jets.  \S\ref{n315sec} discusses in more detail the
giant radio source NGC~315, whose X-ray emission is the most compact
amongst the sample objects.  Appendix~\ref{appendix} gives notes and
comparisons with previous X-ray work for the other seven radio
galaxies of the sample.  Our conclusions are summarized in
\S\ref{consec}.

Throughout the paper we adopt a Friedmann cosmological model with
$H_0$ = 50 km s$^{-1}$ Mpc$^{-1}$, $q_o$ = 0.

\section{Sample Selection and X-ray Exposures} \label{samplesec}

The complete subset of 50 sources from the B2 radio survey identified
with elliptical galaxies brighter than $m_{\rm ph} = 15.7$ (Colla et
al.~1975; Ulrich~1989) is the largest unbiased sample of exclusively
low-power radio galaxies studied with sensitive pointed X-ray
observations to date.  40 of the sources were observed in ROSAT
pointings, half by us through various announcements of opportunity as,
taking into account sources observed by others, we defined sub-samples
with increasingly broader selection criteria so that the largest
possible and least biased subset would have been observed at any one
time.  Following the demise of ROSAT in February 1999, a summary of
the X-ray results for the sample was prepared and appears in Canosa et
al.~(1999), which also references published detailed work by us and
others for some of the individual objects.

Most B2 sample sources were observed with the ROSAT High Resolution
Imager (HRI), but earlier in the mission observations could be
made with the Position Sensitive Proportional Counter (PSPC)
which, while having a poorer spatial response than the HRI, had
spectral capabilities and was more sensitive to faint diffuse
X-ray structures (Briel et al.~1996).

\begin{figure}
\plotone{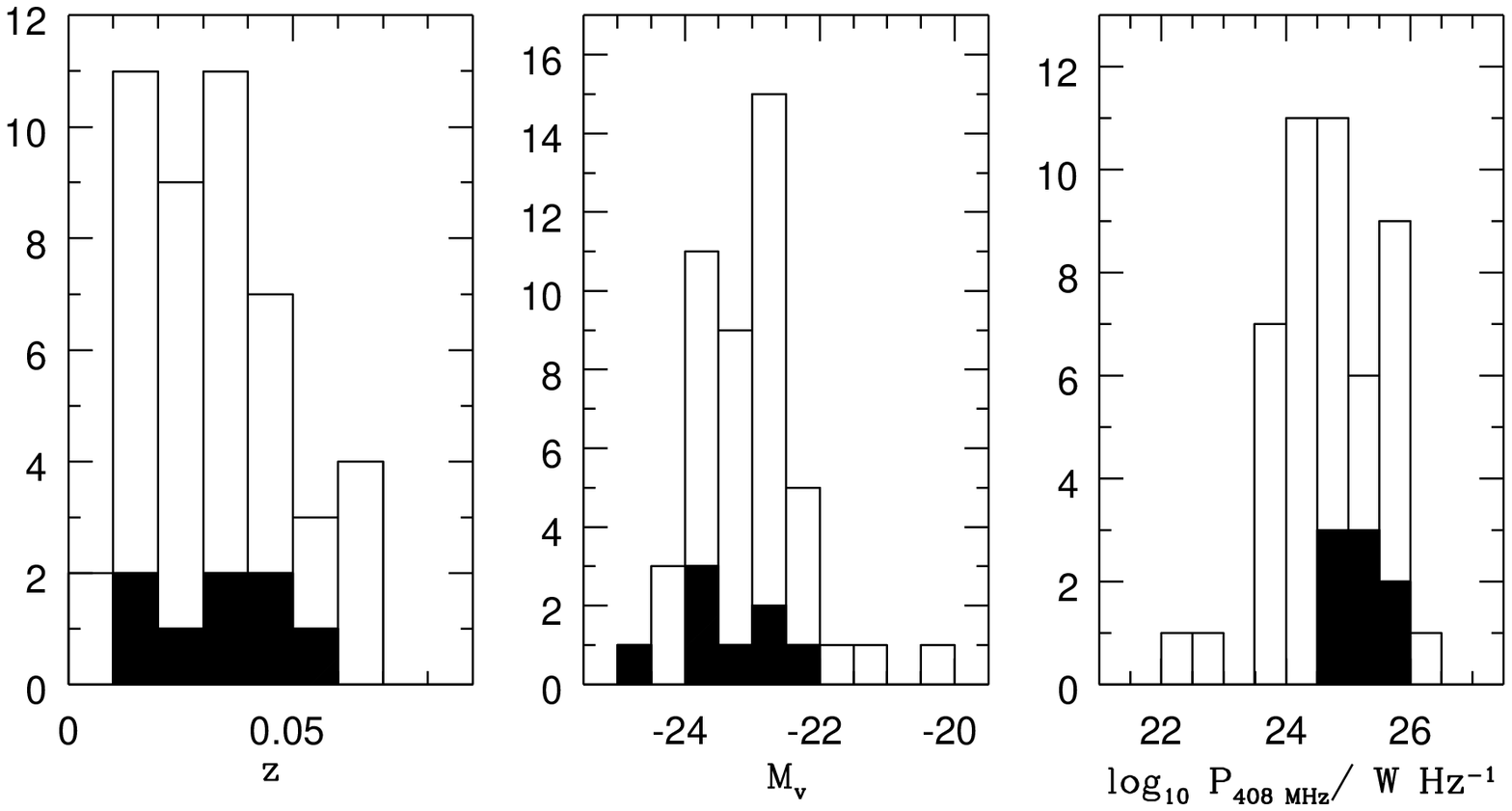}
\caption{ Histograms of redshift, absolute V-band galaxy magnitude
(Ulrich 1989), and 408~MHz radio power (Colla et al. [1975], corrected
to the flux scale of Baars et al.~[1977]) for the complete sample of
47 B2 radio galaxies with $z \leq 0.065$ (Canosa et al.~1999).  Solid
shading shows that the subset of 8 sources in this work are
representative of the complete sample in all three properties except
for total radio power.
\label{histfig}}
\end{figure}

We observed eight sample members (Table~\ref{sampletab}) with the
PSPC, four of which (NGC~315, 4C~35.03, NGC~2484, 3C~449) were also
observed later with the ROSAT HRI by us or others in order to probe in
more detail the nuclear part of the structure.  Figure~\ref{histfig}
plots histograms of redshift, absolute V-band galaxy magnitude, and
408 MHz radio power for the complete sub-sample of the 47 out of 50 B2
sample objects with $z \leq 0.065$, indicating with solid shading the
eight sources discussed in the present paper.  Only in absolute radio
power are the eight observed sources unrepresentative of the sample as
a whole, being biased towards high power: a KS test gives a
probability of only $\sim 3\%$ of the sub-sample being drawn at random
in radio power, whereas similar tests for redshift and galaxy
magnitude give probabilities $> 60\%$.  This bias in total radio power
was unintentional, but most likely occurs through our placement of
sources with better radio data at higher priority for X-ray
observation.  For our observations we also avoided sources in known
clusters (several B2 radio galaxies are in Abell clusters), although
this was a priority for other observers.  Six of the sources had no
previous X-ray observations and two had been observed with the {\it
Einstein\/} Observatory: NGC~315 is reported as unresolved and
variable at $3\sigma$ significance between two observations separated
by about 7 months (\markcite{fab92}Fabbiano, Kim \& Trinchieri~1992),
whereas the emission from 3C~449 was resolved by the Imaging
Proportional Counter (\markcite{mil83}Miley et al.~1983).  Only three
other B2 bright-sample radio galaxies which are not in the Coma
cluster or an Abell cluster have on-axis exposures of $> 5$~ks
with the ROSAT PSPC, and all exhibit extended X-ray emission: 3C~31
(\markcite{tru97}Trussoni et al.~1997; \markcite{kom99}Komossa \&
B\"ohringer 1999), NGC~507 (\markcite{kim95}Kim \& Fabbiano 1995), and
NGC~3665 (\markcite{mass96}Massaglia et al.~1996).



Dates and exposure times for the PSPC observations reported here are
given in Table~\ref{exposuretab}, along with net counts within a
radius large enough to incorporate most of the extended emission.
Earlier reports on some aspects of these data for five of the sources
appear in Worrall \& Birkinshaw (1994), Worrall et al.~(1995), and
Hardcastle et al.~(1998).  A discussion of the extended
emission in 0326+39, 1040+31, and 1855+37
has not appeared elsewhere.  In this paper we present results derived
in a consistent fashion for the gaseous X-ray emitting environments of
all eight sources, and we compare these environments and discuss their
relationships to the radio structures that they host.

The observations of NGC~315 were split into two observing periods
separated by six months.  Because of the reported variability in the
{\it Einstein\/} data, and in order to measure the level of likely
systematic error in our analysis procedures, we treated the two
exposures separately for much of the analysis.  The fluxes measured in
the two exposures are in good agreement (Table~\ref{exposuretab}) and
do not confirm the earlier reports of variability.

 
\section{Spatial distribution of X-ray emitting gas} \label{spatsec}

\begin{figure}
\plotone{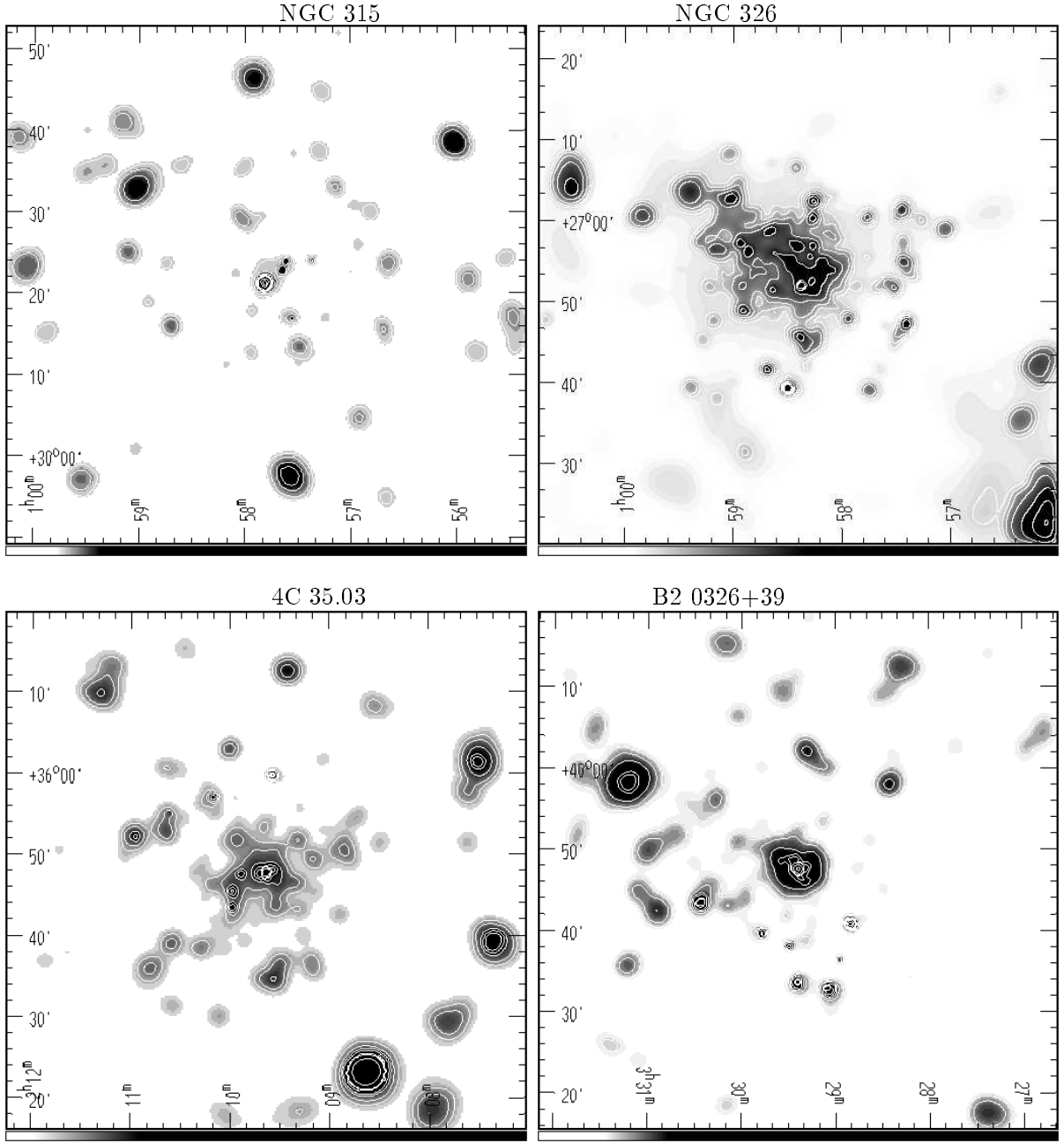}
\caption{
Background-subtracted, exposure corrected, and
adaptively smoothed ROSAT PSPC X-ray images. Coordinates are J2000.
Energy band is 0.4 - 2~keV
(except NGC 326 which is 0.2-2.3~keV).  Contour levels, in units of
$10^{-3}$ cts arcmin$^{-2}$ s$^{-1}$, are as follows.
NGC~315: 0.19, 0.34, 0.6, 1.9, 3.4, 6, 19, 34. NGC~326: 0.18,
0.3, 0.5, 0.83, 1.4, 2.3, 3.9, 6.6. 4C~35.03: 0.19, 0.34, 0.6, 1.1,
1.9, 3.4, 6. B2~0326+39: 0.19, 0.34, 0.6, 1.1, 1.9, 3.4, 6, 11, 19.
\label{contafig}}
\end{figure}

\begin{figure}
\plotone{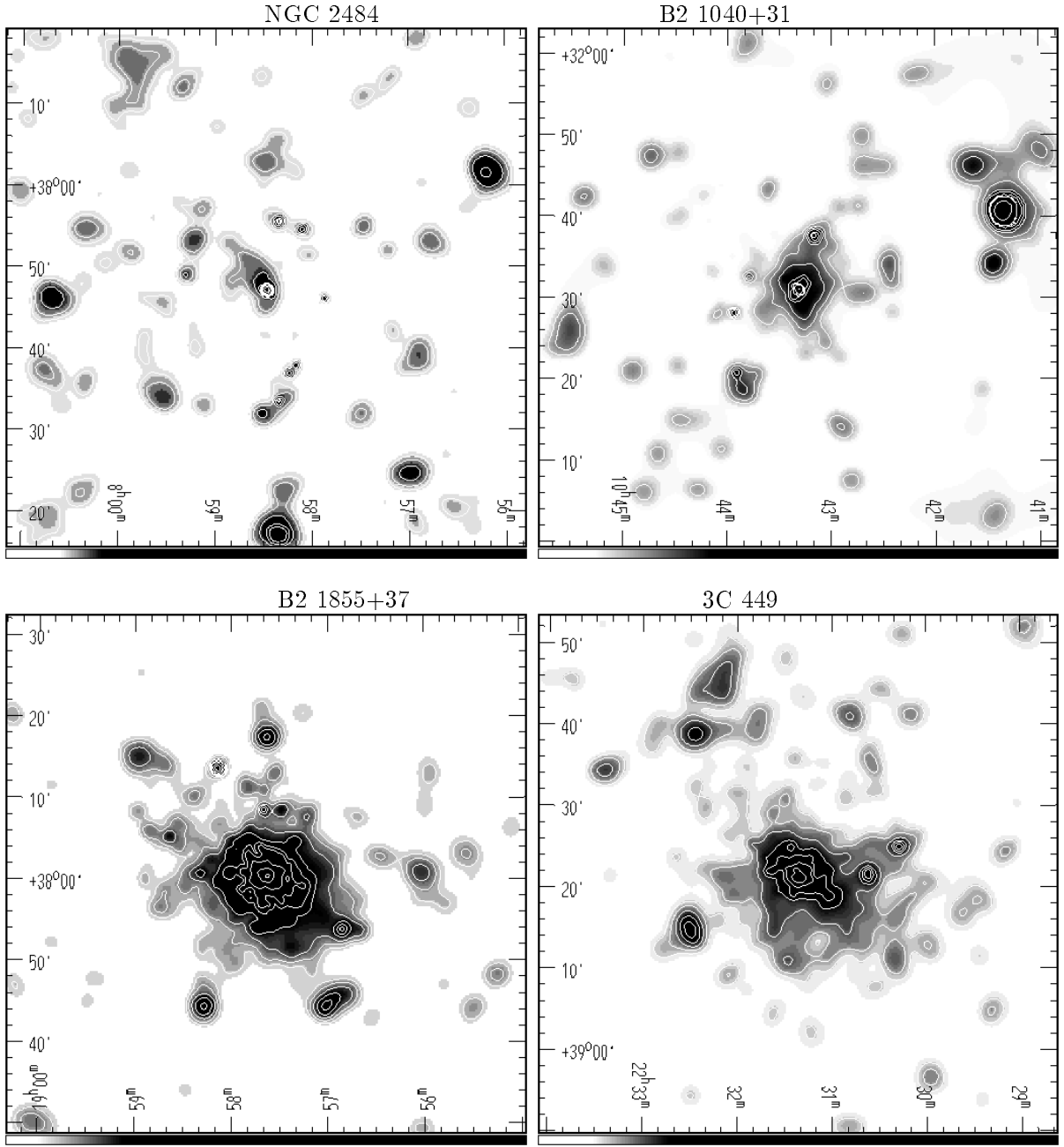}
\caption{
Same as Fig.~\ref{contafig}.
Contour levels, in units of
$10^{-3}$ cts arcmin$^{-2}$ s$^{-1}$, are as follows.
NGC~2484: 0.19, 0.34, 0.6, 1.1, 1.9, 3.4, 6.0, 11.0, 19.0.
B2~1040+31: 0.19, 0.34, 0.6, 1.1, 1.9, 3.4, 6.0, 11.0.
B2~1855+37: 0.19, 0.34, 0.6, 1.1, 1.9, 3.4, 6.0, 11.0, 19.0.
3C~449: 0.19, 0.34, 0.6, 1.1, 1.9, 3.4, 6.0.
\label{contbfig}}
\end{figure}

X-ray images of all eight sources are presented in
Figures~\ref{contafig} and \ref{contbfig}.  For these images the
Extended Source Analysis Software (EXAS) of
\markcite{snow95}Snowden~(1995) was used to model and subtract
background components and correct for exposure and vignetting, except
for NGC~326 which used an earlier procedure which was based on
IRAF/Post Reduction Off-line Software (PROS) tools and which was found
to give similar results (\markcite{wbc95}Worrall et al.~1995).  The
adaptive smoothing technique described by \markcite{wbc95}Worrall et
al.~(1995) has been applied to all the images in order that the
display should distinguish clearly between unresolved sources and
diffuse emission.  The PSPC point response function (PRF) increases at
off-axis angles, as is evident in the images.  Extended emission is
seen around all the radio galaxies, which lie in the center of the
fields.  The angular extent is not anti-correlated with luminosity
distance, as would be expected if the radio galaxies are in
environments of similar spatial scale.



Although asymmetries are evident in the gas distributions, the
construction of a radial profile, after subtraction of emission from
contaminating unresolved sources, is a useful tool for characterizing
with a small number of parameters the spatial distribution of gas.  It
is a reasonably robust way of separating unresolved
emission associated with the radio galaxy, and the models can be
extrapolated to radii where emission no longer lies significantly above
the level of the background in order to get the best possible
measurement of the total luminosity of the source.  In comparison with
\markcite{worr94}Worrall \& Birkinshaw (1994) and
\markcite{can99}Canosa et al.~(1999) the on-source extractions used in
our radial profiles are generally to larger radii, to allow a better
description of the fall-off of gas with radius and a more precise
determination of gas temperature. The fitting is therefore less good
at characterizing the properties of compact, unresolved, emission.
Several faint nearby contaminating unresolved sources were defined by
eye as localized excesses, and regions of typically 1 to 2.5 arcmin in
radius enclosing them were excluded from the on-source and background
areas for both structural and spectral analyses.

\begin{figure}
\plotone{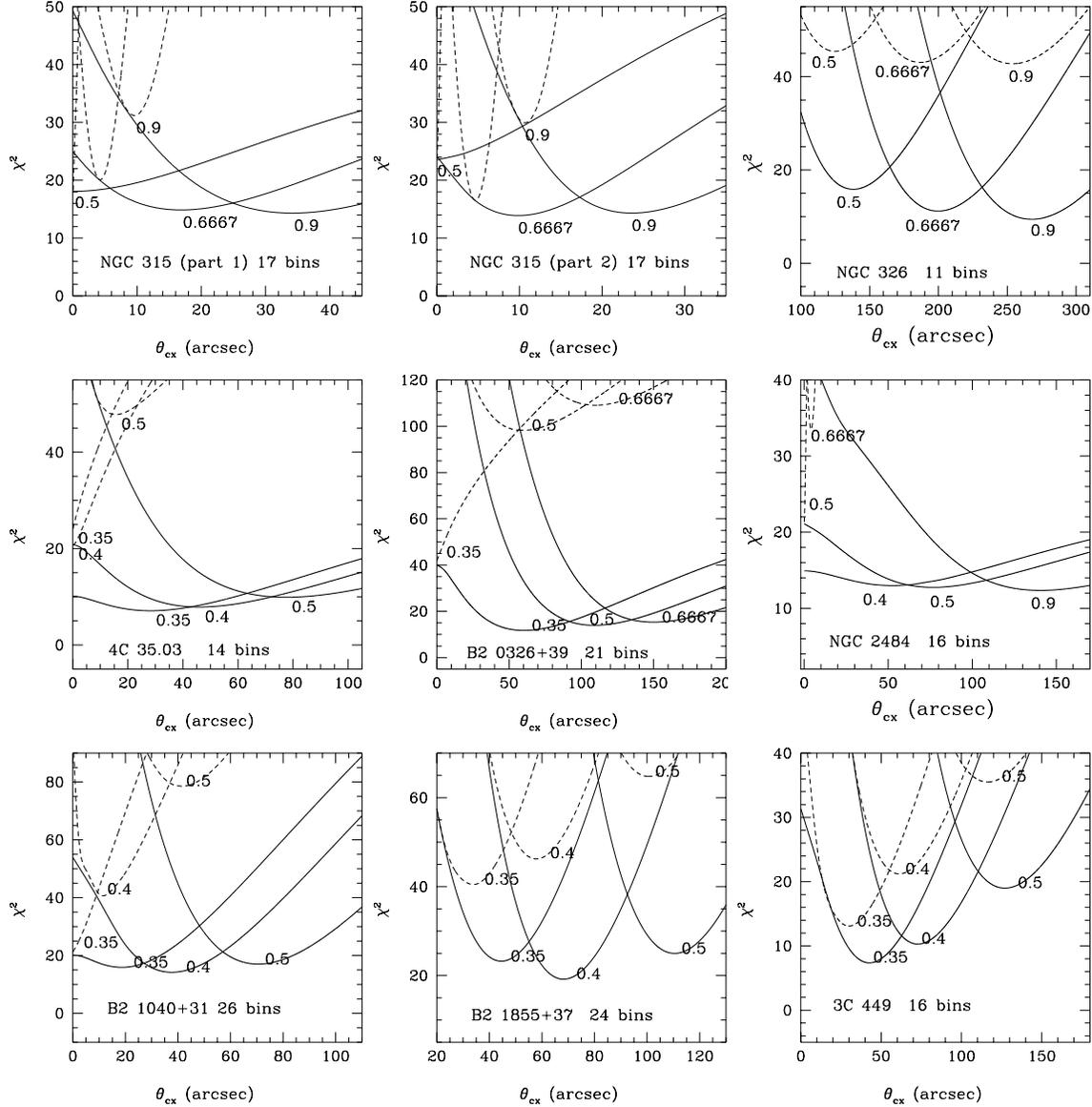}
\caption{
$\chi^2$
versus core radius of the $\beta$ model for fits to a $\beta$ model
alone (dashed curves) and the combination of a $\beta$ model and
unresolved component (solid curves).  For each source and both models,
fits were run for $\beta = 0.35$, $0.4$, $0.5$, $0.6667$, and $0.9$.
We show only a selection of $\beta$ values, always including those
giving a minimum $\chi^2$ for each model, in order to simplify the
presentation. Two separate panels for NGC~315 correspond to the
January and July 1992 observations (see \S\ref{samplesec}).
\label{chifig}}
\end{figure}

\begin{figure}
\plotone{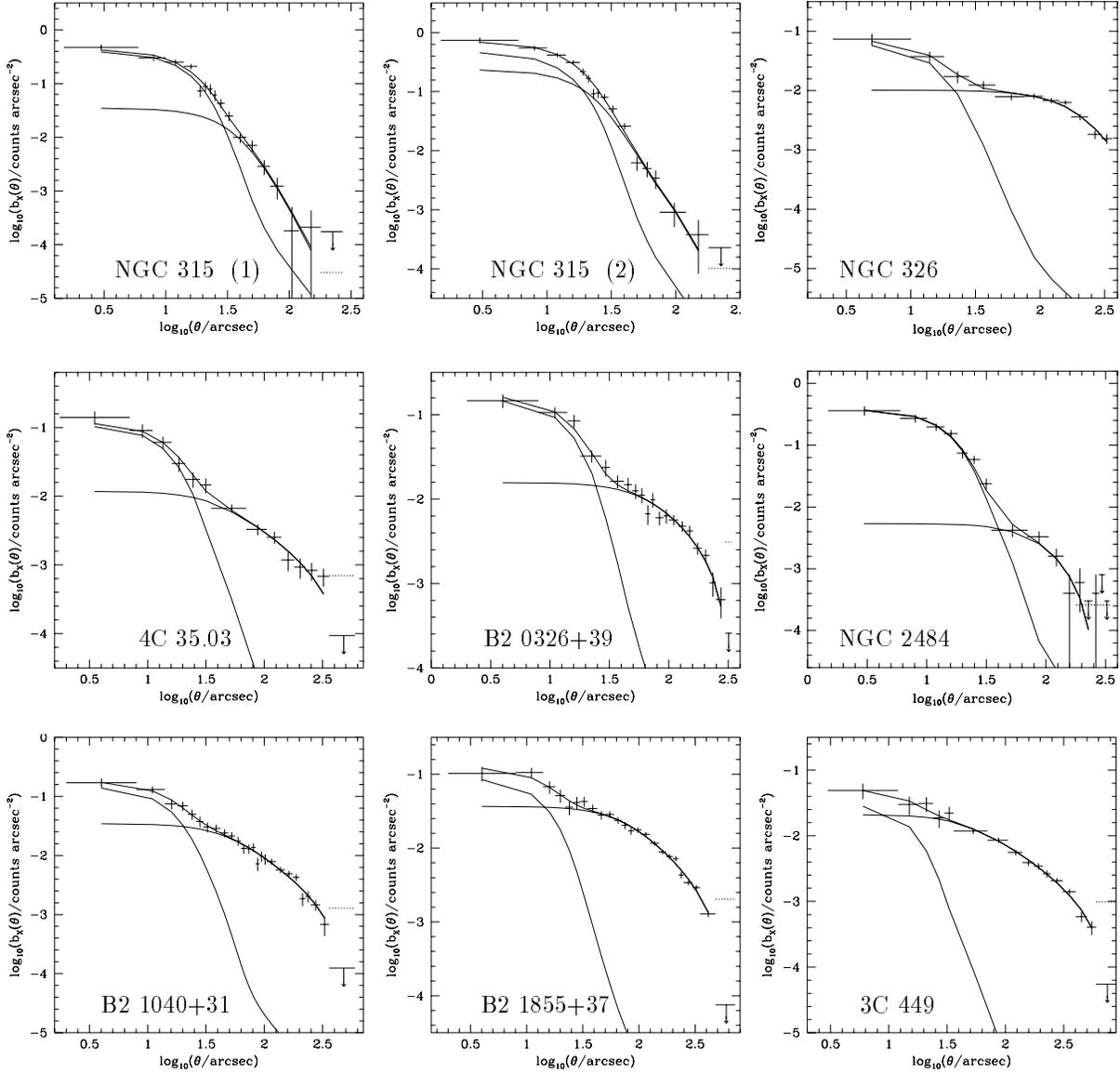}
\caption{
Background-subtracted radial profiles and best fits
to the combination of a $\beta$ model and unresolved component. Dotted
lines show contributions of the model to the background region, as
taken into account in the fitting, for all but NGC~326 whose
background region was beyond the radius shown.
\label{radfig}}
\end{figure}

Single component $\beta$-models\footnote{Counts per unit area per unit
time described by $B_o (1 + {\theta^2 \over \theta_{\rm cx}^2})^{0.5 -
3 \beta}$}, describing gas in hydrostatic equilibrium (e.g.,
\markcite{sar86}Sarazin~1986), and a combination of $\beta$-model and
unresolved emission, have been convolved with the energy-dependent PRF
and fitted to the radial profile for each source.  Only counts within
the energy band for which the PRF is well modelled (0.2-1.9 keV; Briel
et al.~1996) are used in the spatial analysis.  In contrast to
\markcite{worr94}Worrall \& Birkinshaw (1994) who assumed $\beta =
2/3$, fits were performed for a range of $\beta$ between 0.35 and 0.9.
In general results are fairly insensitive to $\beta$, but the values
of $\beta$ and core radius, $\theta_{\rm cx}$, are highly correlated,
as shown in Figure~\ref{chifig}, making FWHM a better indicator of
spatial extent than $\theta_{\rm cx}$.  Radial profiles and the
best-fit models are given in Figure~\ref{radfig}, and parameter values
are listed in Table~\ref{radialtab}.  All profiles are significantly
improved by the inclusion of nuclear unresolved emission, except perhaps for
NGC~315.  In most sources the fraction of counts in this unresolved
emission is relatively small, largely due to a large on-source
extraction radius.




4C~35.03, NGC~2484, and 3C~449 were also observed with the HRI (see
notes in Appendix~A). In each case the amount of nuclear unresolved
flux as measured with the PSPC matches that measured with the HRI, and
so we can conclude that the compact component is more peaked and of
angular size a factor of 3--4 times smaller than the PSPC PRF shown in
Figure~\ref{radfig}.  The more complicated case of NGC~315 is
discussed in \S\ref{n315sec}.

\section{The X-ray spectra} \label{specsec}

Spectral fits to the data were carried out over the full energy range
of the PSPC, and results are given in Table~\ref{spectraltab}.  For
sources where the spatial fitting gives $\geq 14\%$ of the counts in a
central component we have attempted a two-component (thermal plus
power-law) spectral fit. The relative numbers of counts in the two
components are generally less well constrained than in the
radial-profile component separation; results in column~5 of
Table~\ref{spectraltab} (expressed only over the energy band 0.2-1.9
keV for consistency with the spatial separation) can be compared with
column~2 of Table~\ref{radialtab}.  Best-estimate fluxes, luminosities
(Table~\ref{spectraltab}) and temperatures (Table~\ref{coolingtab})
for the thermal gas are interpolated from the single-component and
two-component fits.  Spectral fits assume the only absorption to be
that due to gas in the line of sight in our Galaxy, and we have fixed
the abundances at 30\% solar, consistent with results for more
luminous clusters.  The abundance fraction is fairly poorly
constrained by the PSPC data, particularly for two-component fits.
However, for all sources the best fit is for $\leq 40$\% solar, and
fits with 100\% solar abundances give values of $\chi^2$ more than 2.7
larger than $\chi^2_{\rm min}$, indicating they are unacceptable at $>
90$\% confidence. Although the PSPC's spectral resolution is
relatively poor, the peak response is well matched to the
temperatures found for the gaseous distributions around these
radio galaxies.



\section{Characteristics of the Radio-Galaxy Environments}
\label{physsec}

\begin{figure}
\plotone{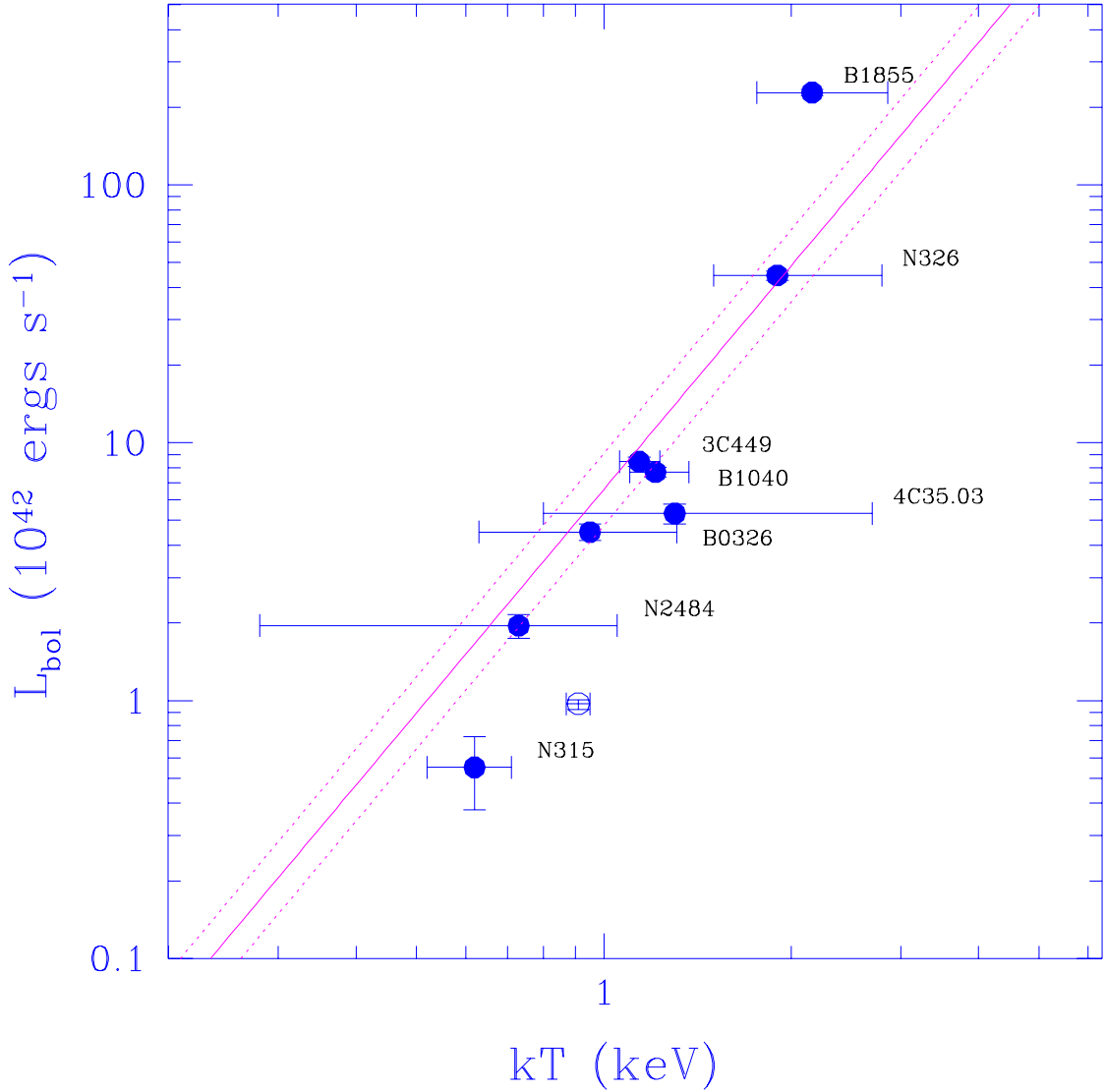}
\caption{ Luminosity (from Table~\ref{spectraltab} but with bolometric
corrections applied) vs. temperature (from Table~\ref{coolingtab}). 
The open circle is the result for NGC 315 assuming no power-law component.
The solid and
dashed lines are the best-fit relation and rms deviations for more
luminous clusters ($\sim 10^{44} - 10^{46}$ ergs s$^{-1}$) from
Arnaud \& Evrard (1999). \label{ltfig}}
\end{figure}

In Figure~\ref{ltfig} we compare the bolometric luminosity and
temperature of the X-ray emitting gas around each radio galaxy to the
extrapolated but well constrained luminosity-temperature ($L_{\rm bol}
- kT$) relation for more luminous clusters (with $kT > 2$ keV) from
Arnaud \& Evrard (1999).  Agreement is good.  Since X-ray luminosity
is governed by the gas mass while the temperature is determined by the
total gravitating mass, the result implies that there is no
relationship between the presence of a radio galaxy and the gas
fraction of the environment.  This is an interesting point, since the
X-ray gas may play a role in the fuelling of radio galaxies,
particularly in the presence of cooling flows.
However, as seen from Table~\ref{coolingtab}, NGC~315 is the only source
with a central cooling time which is much less than
the Hubble time, and this is the only source for which
the gas distribution is of galaxy size rather than
cluster (or group) size.  There is no evidence for cluster-scale
cooling flows in the present data.

\begin{figure}
\plotone{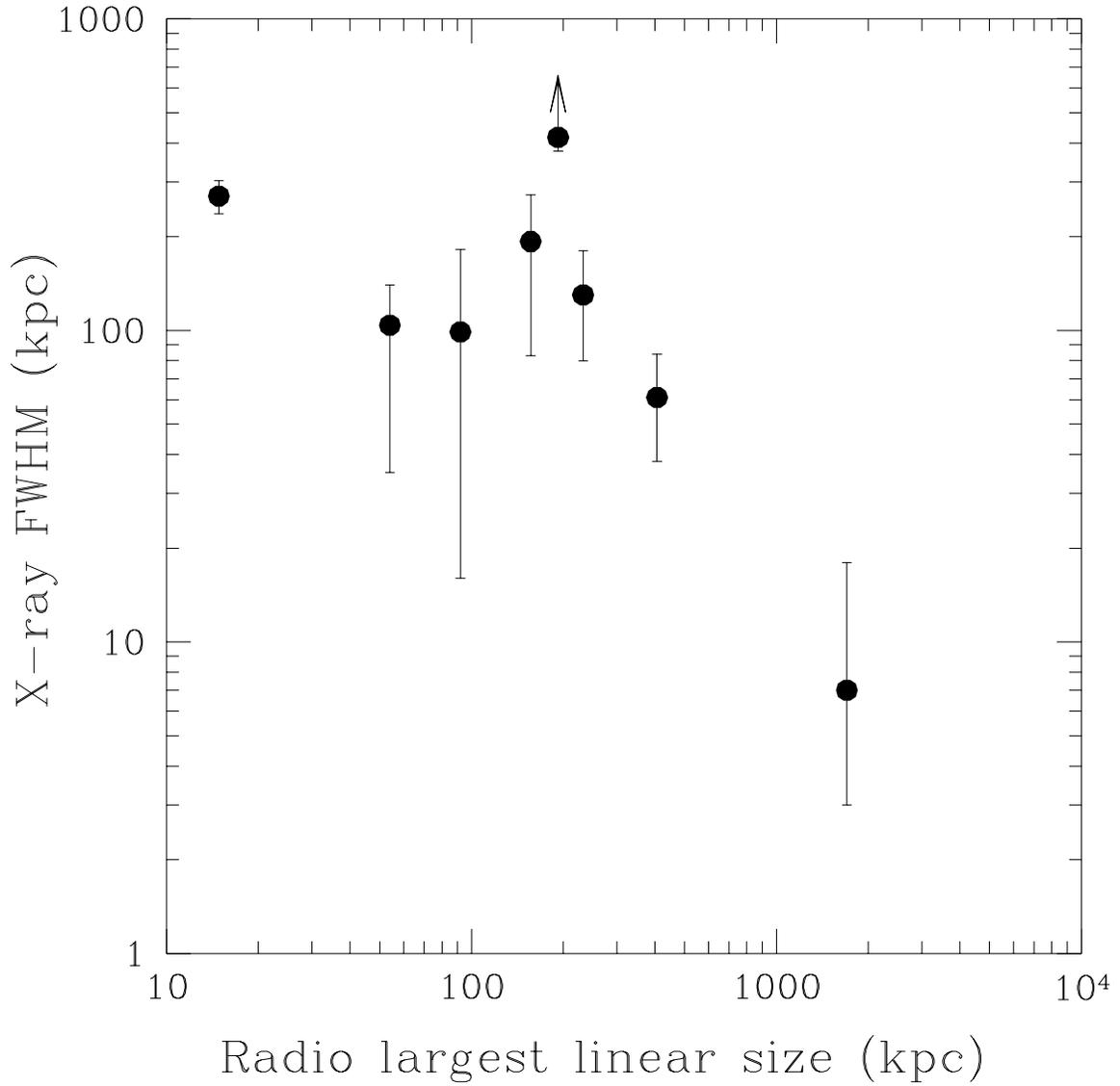}
\caption{ The largest linear size of radio structure
(Table~\ref{sampletab}) is uncorrelated with
the linear size of the X-ray emitting atmosphere
(Table~\ref{radialtab}).  \label{lasfig}}
\end{figure}

Radio-source size is not correlated with the size of the X-ray
emitting medium (Figure~\ref{lasfig}), where note that FWHM
(Table~\ref{radialtab}) is a scale factor for the size of the X-ray
emitting medium rather than its total (much larger) extent.  The lack
of a correlation is perhaps not surprising, given that the sound
crossing time in X-ray gas of extent $10 - 100$~kpc, at $\sim (20 -
200)~{(kT/{\rm keV})}^{-1/2}$~Myr, is comparable to typical radiative
ages of low-power radio galaxies, at $10 - 100$~Myr
(\markcite{parm99}Parma et al.~1999), and so only for the oldest
sources might it be expected that the gas has had time to adjust to
the presence of the radio galaxy.  Conversely, the lack of correlation
suggests that it is small-scale processes, on size scales less than
those of the overall gaseous environments, which are the major
influence on radio-source dynamics and propagation.

\begin{figure}
\plotone{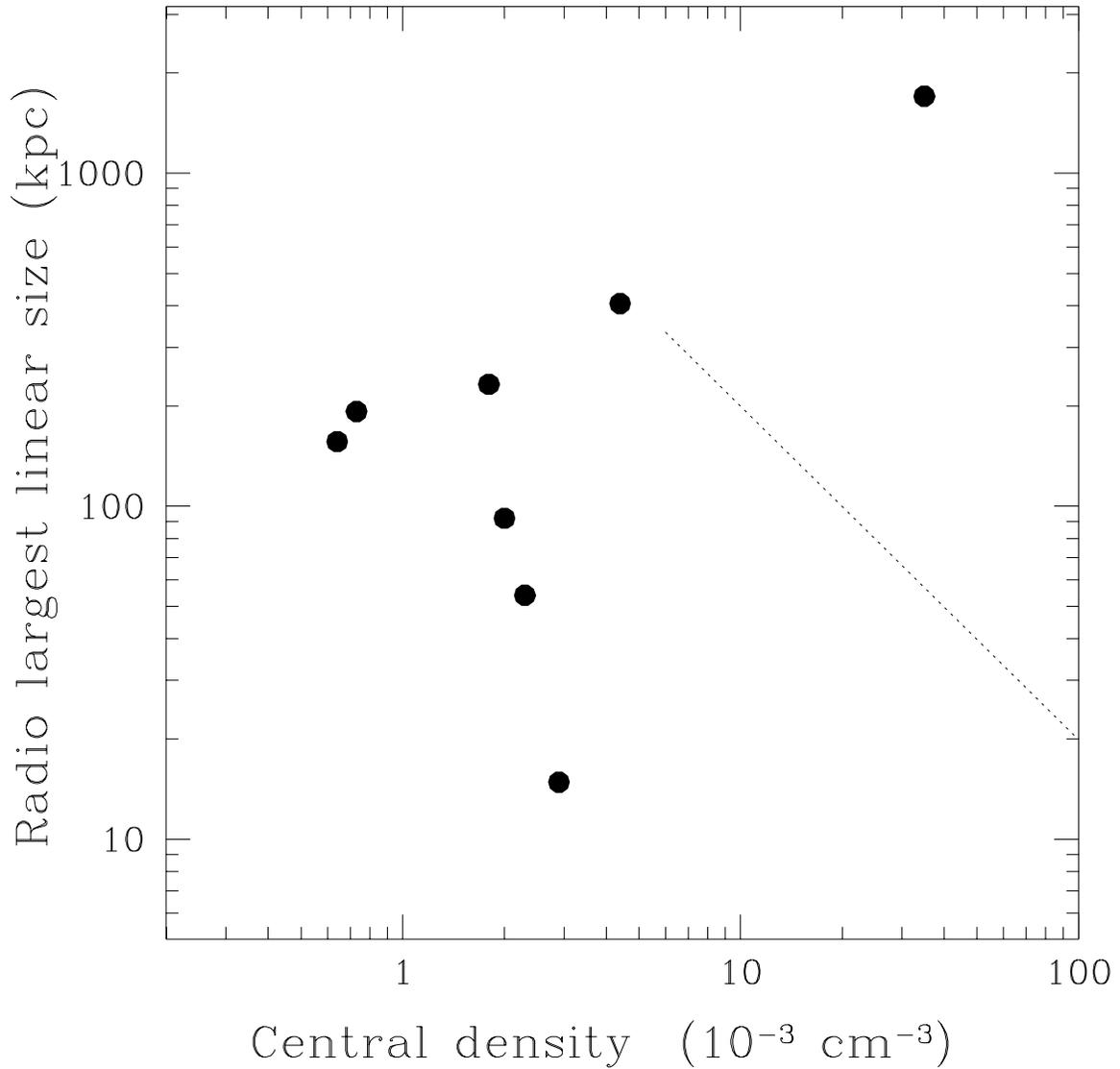}
\caption{ The largest linear size of radio structure
(Table~\ref{sampletab}) is uncorrelated with
the central density of X-ray emitting gas
(Table~\ref{coolingtab}).  Dashed line is 
Morganti et al.~(1988)'s
correlation for an inhomogeneous sample of low-luminosity radio
galaxies based on {\it Einstein\/} data. \label{morgfig}}
\end{figure}

As the spatial resolution of X-ray measurements improves, it is not
surprising that inferred central gas densities decrease, as gas is
resolved into larger regions.  \markcite{morg88}Morganti et al.~(1988)
infer significantly higher central densities than our values
(Table~\ref{coolingtab}) for a sub-sample of B2 radio galaxies
measured with {\it Einstein\/}, although none of the objects is in
common with sources in this work.  Moreover, they tentatively claim an
anti-correlation between largest linear extent of radio source and
central gas density, interpreting this to be evidence that the gas has
a direct influence on the morphology of the radio source.  No such
anti-correlation is evident in our results (Figure~\ref{morgfig}), and
again this can be understood in terms of the long timescale for X-ray
gas to react to the influence of a radio source.


For all the sources we have compared minimum pressures in various
parts of the radio structures with the pressure from the X-ray gas.
For the radio pressure calculations it was assumed that the
synchrotron spectrum extends above 50~MHz, that the electron energy
spectrum is of $E^{-2}$ form, and that electrons and protons
contribute equally to the internal energy.  Results for six of the
sources are given in Figure~\ref{pressfig}, and individual notes
appear in the Appendix~\ref{appendix} and \S\ref{n315sec}.


Within the limitations of the available radio-mapping data, the
diffuse outer parts of all eight radio sources exhibit minimum
pressures close to or below the local pressures implied by the
best-fit X-ray atmospheres. In inner parts of the sources, where
strong jets are seen, the minimum pressures in the kpc-scale
structures sometimes lie above the
pressure of the ambient gas. Thus if we are to believe that the radio
structures are in pressure equilibrium with the external medium, we
require that at small angles from the cores there be an additional
confinement mechanism, while at large angles from the cores additional
internal pressure is needed.  There are several mechanisms for
additional internal or external pressure.  By definition, the minimum
pressures are likely to be underestimates, with higher true pressures
likely if the sources have substructure, contain a large population of
non-radiating relativistic particles or entrained ambient material (on
which limits can be placed by depolarization studies if the magnetic
field has a simple topology), or are simply far from equipartition. If
the sources are not in the plane of the sky, then projection effects
tend to overestimate the local ambient pressures. Additional external
confinement may be provided by magnetic fields or local pressure
enhancements near the radio sources and associated with the source
dynamics.

There are three exceptions to this general pattern.  3C~449 and
4C~35.03 are underpressured over their entire structures, and NGC~315
may remain close to pressure equilibrium throughout its length.  NGC~315
is 1.7~Mpc in size, easily the largest in our sample, and here an
extrapolation beyond the region of clear X-ray detection suggests the
source may remain close to pressure equilibrium over a factor $1600$
variation in the ambient gas pressure. We would speculate that it is
\textit{because} NGC~315 is close to pressure equilibrium that it is
able to grow to such a large scale. The detected X-ray emitting
atmosphere extends only to $\sim 2.5$~arcmin (70~kpc), however. Thus it
is not certain that the extrapolated X-ray pressure estimates at the
largest scale of the source are realistic.

\section{The Special Case of NGC 315} \label{n315sec}

The spatial structure of the X-ray emission around NGC~315 is unique
amongst the sample in its compactness.
\markcite{can99}Canosa et al.~(1999) attributed all the (mildly
extended) ROSAT HRI flux, within an on-source extraction radius of
50~arcsec, to unresolved emission of luminosity $(1.3 \pm 0.1) \times
10^{42}$ ergs s$^{-1}$ (statistical errors), because of limitations
due to the aspect-correction errors in the ROSAT Standard Analysis
Software System (SASS) data.  However, the PSPC data show that
extended emission is definitely present (Fig.~\ref{radfig}), and
the nature of the most compact regions of the source need
further investigation.

Although the PSPC spatial analysis prefers the presence of a point
source with only marginal significance (and not at all when the two
observations are combined), a substantial AGN-related
contribution to the PSPC data
seems justified since

\begin{enumerate}
\item  there is a significant
improvement in spectral fits when a power-law component is included
(Table~\ref{spectraltab}), 

\item a two-temperature thermal fit 
(assuming 30\% cosmic abundances) gives 56\% of
the counts in a thermal of temperature 0.6~keV, and the rest in a hot
component of 2.8~keV, in which case the temperature is surprisingly high
given the emission's compactness and luminosity (Fig.~\ref{ltfig}), and

\item the presence of significant
power-law emission is suggested by the relatively strong radio core
and the correlation of radio-core and compact-X-ray luminosity shown
by the larger sample of B2 radio galaxies, including NGC~315
(\markcite{can99}Canosa et al.~1999).  

\end{enumerate}

\noindent
However, the luminosity in unresolved emission from the PSPC spectral
(and spatial) fitting, at $\sim 0.5 \times 10^{42}$ ergs s$^{-1}$, is
significantly smaller than the total luminosity in the HRI data, and
roughly equal to the contribution from gas measured with the PSPC.
This suggests that the unresolved emission increased by a factor up to
two (depending on the precise contribution of thermal emission to the
HRI flux) in the two years separating the PSPC and HRI observations.
This is not impossible if the emission is non-thermal and
radio-related in nature, and the result is supported by the earlier
{\it Einstein\/} report of variability.

We have re-examined the HRI data for spatial extent after first
applying software just released to correct approximately the effects
of a recently identified programming error in the SASS
aspect-correction software (\markcite{harris99}Harris~1999).  The
result is that the HRI data still give an unacceptable fit to the
nominal PRF, and prefer a single-component $\beta$-model, but of very
small core radius ($\theta_{\rm cx}=3$~arcsec, $\beta = 2/3$),
inconsistent with the PSPC data.  The extent of residual aspect errors
is uncertain, but broadening the PRF by a few arcsec results in a
combination of a point source and beta component being the preferred
model, and brings results into closer agreement with those from the
PSPC data.

A further complication is that this is the only source from the
current sample where the cooling time is sufficiently short
(Table~\ref{coolingtab}) to suggest the presence of a cooling flow.
However, although the cooling time is estimated to equal the Hubble
time at a radius of $\sim 20$~kpc, there is no abrupt steepening in
the PSPC radial profile at such a radius (Fig.~\ref{radfig}) as might
be expected from the on-set of a cooling flow.  If for the HRI data a
single-component cooling-flow model fit (of the type described in
\markcite{hwb99}Hardcastle, Worrall \& Birkinshaw 1999) is attempted,
then no improvement over a point source plus $\beta$-model is achieved
unless the cooling radius of the gas is $\lesssim 5$~kpc, which not
only implies an atmosphere less than $\sim 300$ Myr old, but also is
inconsistent with the larger cooling radius suggested by the PSPC
data.  It certainly remains possible that a very small-scale cooling
flow contributes to the unresolved emission, and we now await data
from the {\it Chandra\/} Observatory, whose $\sim 0.5$~arcsec
resolution will significantly improve our knowledge of the inner
regions of this source.

The minimum pressures of various regions in the inner radio jet of
NGC~315 were estimated from a 1.4~GHz WSRT map kindly provided by Alan
Willis, and supplemented by the lower-resolution results of
\markcite{mack98}Mack et al.~(1998) in the outer parts of the
source. Figure~\ref{pressfig} shows these minimum pressures
superimposed on the pressure of the X-ray emitting atmosphere
estimated from the best-fitting $\beta$-model (Table~\ref{coolingtab})
and extrapolated beyond the regions of clear X-ray detection. This
figure suggests that if the radio jet lies close to the plane of the
sky, then only the knot at about 5~arcmin from the core may lie
significantly out of pressure balance: all the other parts of the jet
and the outer radio source may lie near pressure equilibrium or be
confined by the external medium. Given the wide range in pressures (a
factor $1600$), this suggests that the structure of NGC~315 may be
strongly influenced by the X-ray emitting atmosphere over the entire
region outside 20~arcsec from the nucleus. Within 20~arcsec of the
nucleus the radio data available to us have insufficient resolution to
allow a useful estimate of the minimum pressure: similarly, structures
of smaller angular size in other parts of NGC~315 may be far from
pressure balance.

Figure~\ref{n315fig} shows an overlay of the inner part of the radio
source on the PSPC X-ray emission.  There is a possible small-scale
extension of the X-ray emission along both jets, but the two X-ray
excesses to the northwest, one of which lies on the jet, are most
likely associated with background sources for which optical candidates
are visible on the Palomar Sky Survey images.


\section{Conclusions} \label{consec}

The analyses of these eight B2 radio galaxies lead to
several conclusions which we may expect to be generic of low-power
radio galaxies as a class:

\begin{enumerate}

\item All eight X-ray sources exhibit multiple components.

\item The extended atmospheres have a wide range of linear sizes, and
follow the cluster X-ray luminosity/temperature correlation.

\item No large-scale cooling flows are found, although a small-scale
cooling flow may be present in NGC~315.

\item There is no correlation of radio-source size with the scale or
density of the X-ray atmosphere, suggesting that it is processes on
scales less than those of the overall gaseous environments which are
the major influence on radio-source dynamics.

\item The outer parts of the radio sources are usually
pressure-confined by the X-ray atmospheres, if they are close to
minimum energy.

\item For NGC~315, an extrapolation of the pressure of the atmosphere
suggests that pressure balance may be maintained over a factor of
$\sim 1600$ in gas pressure and an overall size of $\sim 1.5$~Mpc,
although the detected X-ray gas has a FWHM of only $\sim 5$~kpc.
Deeper X-ray images are needed to test this possibility.

\end{enumerate}

\acknowledgments

We thank Alan Willis for providing the 1.4~GHz radio map of NGC~315,
and Martin Hardcastle for discussions and software used to calculate
minimum pressures in the radio sources.  We acknowledge support from
NASA grant NAG~5-1882.

\appendix

\section{Notes on individual Sources} \label{appendix}

\noindent{\bf NGC 315}.  This source is discussed in detail in
section~\ref{n315sec}.

\begin{figure}
\epsscale{0.75}
\plotone{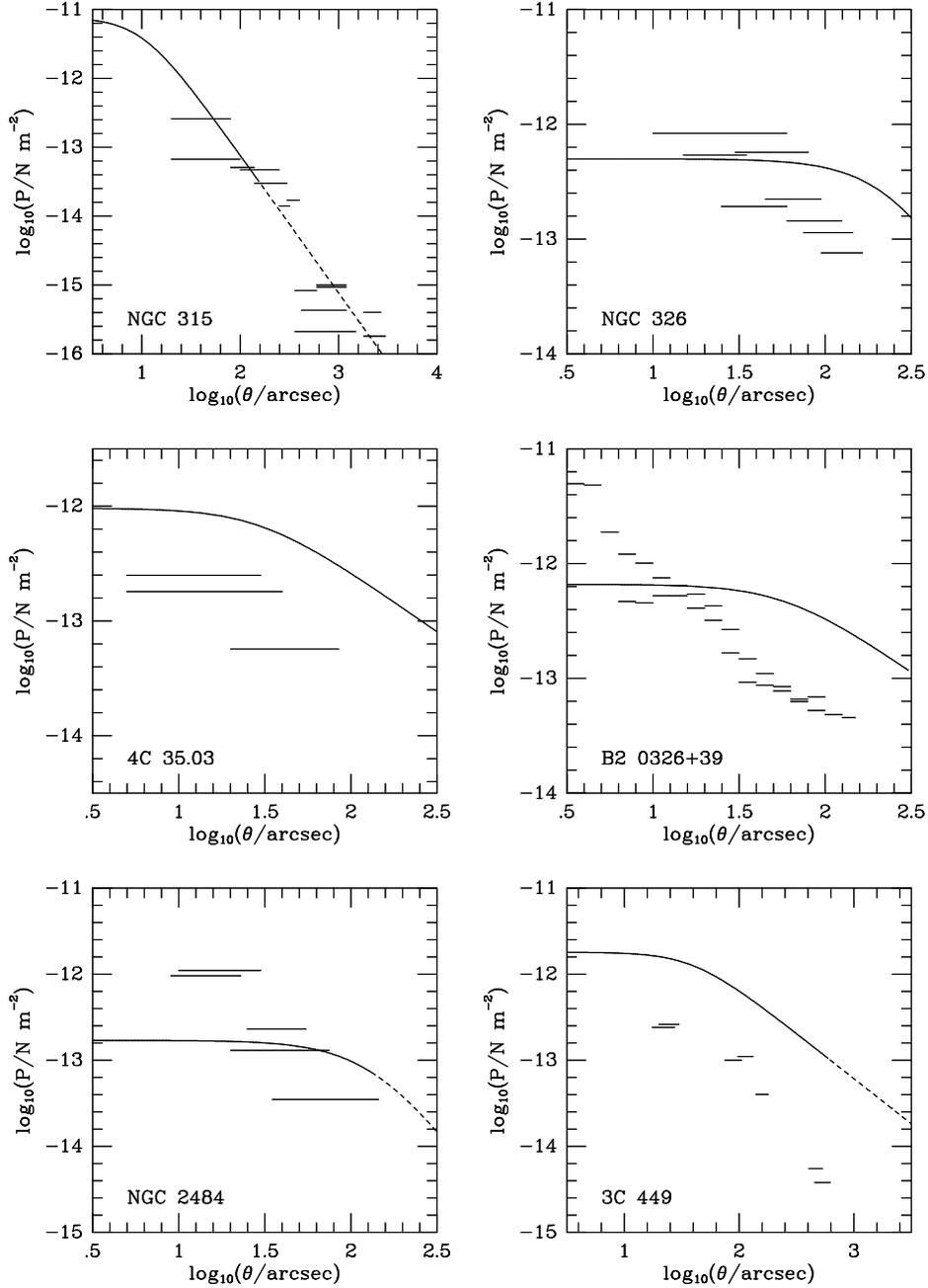}
\caption{ The thermal pressures in the atmosphere of NGC~315, NGC~326,
4C~35.03, B2~0326+39, NGC~2484, and 3C~449 as deduced from fits to the
X-ray images (solid line, shown dashed where extrapolated beyond
region of clear X-ray detection), compared with minimum internal pressure
estimates (the horizontal bars, which indicate the range of angles
over which the pressure estimates apply). The internal pressures are
based on maps from Willis \& O'Dea (1995, private communication) and
estimates from Mack et al.~(1998) for NGC~315; 1.4- and 4.9-GHz maps
from Worrall et al.~(1995) for NGC~326; pressures in Parma et
al.~(1986) for 4C~35.03; pressures in Bridle et al.~(1991) for
B2~0326+39; an unpublished 4.9-GHz map from Birkinshaw \& Davies for
NGC~2484; and pressures in Hardcastle et al.~(1998) for 3C~449.
\label{pressfig}}
\end{figure}

\begin{figure}
\epsscale{1.0}
\plotone{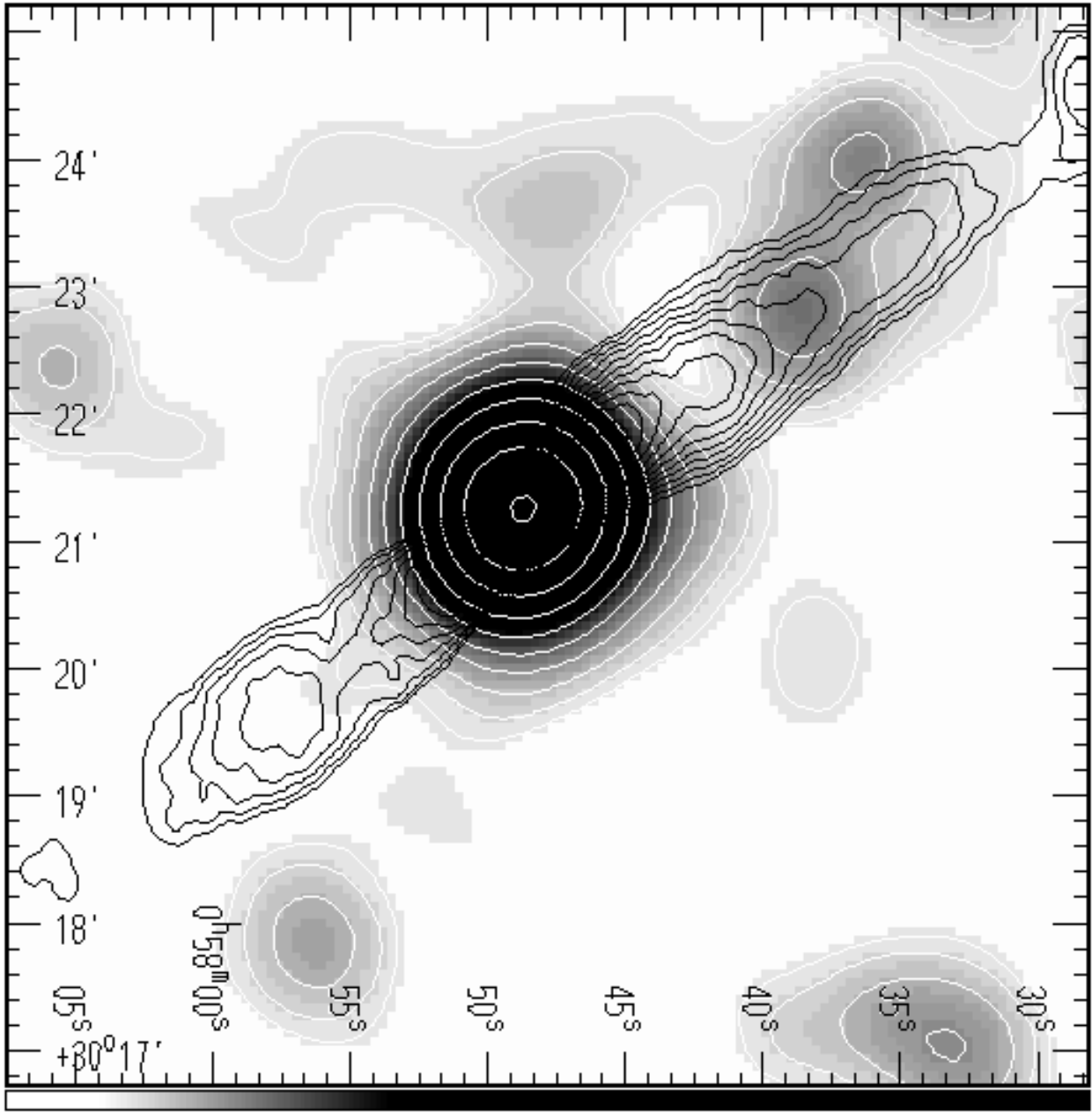}
\caption{ Contours show inner part of a 1.4~GHz radio map taken with
the DRAO Seven-antenna Synthesis Telescope by Willis \& O'Dea (1995,
private communication). The ROSAT PSPC X-ray image shown in
grey scale is smoothed with a Gaussian of $\sigma = 20$~arcsec.
\label{n315fig}}
\end{figure}

\noindent{\bf NGC 326}. Further results based on the ROSAT PSPC
observation appear in \markcite{wbc95}Worrall et al.~(1995), including
an interpretation of the strongly distorted radio structure as due to
a buoyant backflow of radio plasma in the X-ray emitting medium.
\markcite{wern99}Werner, Worrall \& Birkinshaw (1999)'s spectroscopy
of cluster galaxies gives a velocity dispersion which is consistent
with expectations from the X-ray derived cluster properties.
The minimum radio pressures presented in Figure~\ref{pressfig}
were estimated
from the low-resolution 1.4~GHz radio map and higher-resolution
4.9-GHz maps described by
\markcite{wbc95}Worrall et al.~(1995). The inner radio jets
appear to be close to pressure balance with the X-ray gas, while the
outer parts of the source appear underpressured, in accordance with
the usual pattern (\S\ref{physsec}).

\noindent{\bf 4C~35.03}. The ROSAT HRI data are consistent with an
unresolved source of luminosity $10^{42}$ ergs s$^{-1}$
(\markcite{can99}Canosa et al.~1999), which is about 25\% that of the
thermal gas (Table~\ref{spectraltab}) and in good agreement with the
fraction of unresolved emission in the PSPC radial profile
(Table~\ref{radialtab}).  The radial profile is relatively insensitive
to the value of $\beta$, but prefers a flatter distribution than
$\beta = 2/3$ which was assumed by \markcite{worr94}Worrall \&
Birkinshaw (1994).  The unresolved X-ray emission appears not to be a
smooth continuation of the extended gas: it may predominantly be
either a galaxy-scale atmosphere (\markcite{tru97}Trussoni et
al.~1997) in which case it would be cooling rapidly, or a point-source
component associated with the active nucleus.  The latter
interpretation is favored here.  Our approved {\it Chandra\/}
observation should settle this matter.  The radio appearance of the
source has been described by \markcite{fant86}Fanti et al.~(1986): it
exhibits small ($\approx 20$~arcsec) jets to either side of the core,
embedded in a roughly elliptical halo of lower-brightness
emission. The minimum radio pressures in the jets and the halo, as
calculated by \markcite{parm86}Parma et al.~(1986), are a factor of four or
more below the pressure of the X-ray emitting atmosphere, indicating
that the source is likely to be well-confined by the external
gas. This may account for the sharp edges of the
radio image, and suggests that even the inner radio jets' dynamics are
strongly affected by the gas environment.

\noindent{\bf 0326+39}. In \markcite{can99}Canosa et al.~(1999) our
spatial decomposition of the PSPC data uses a smaller on-source region
to probe the unresolved X-ray component.  There are no ROSAT HRI
observations of this source.  The source was mapped in detail at radio
frequencies by \markcite{brid91}Bridle et al.~(1991), and the minimum
pressures in the jets shown in Figure~\ref{pressfig} are taken from
Figure~14 in that paper (adjusted to our choice of $H_0$). It can be
seen that the jets are overpressured relative to the X-ray gas within
about 20~arcsec of the core, and underpressured thereafter. This angle
from the core marks an abrupt transition in the radio image, where the
jets decrease suddenly in brightness. It can be seen that this does
not correspond to any particular feature in the best-fitting (and
flat) $\beta$-model atmosphere, but higher resolution and more
sensitive X-ray observations are required.

\noindent{\bf NGC 2484}. The ROSAT HRI data are consistent with an
unresolved source of luminosity $(3.7 \pm 0.4) \times 10^{42}$ ergs
s$^{-1}$ (\markcite{can99}Canosa et al.~1999).  The best-fit
luminosity in unresolved emission from the PSPC is $3.6 \times
10^{42}$ ergs s$^{-1}$, in excellent agreement.  Our forthcoming {\it
Chandra\/} observation should discriminate between an AGN and
compact-gas origin for the unresolved emission.  The PSPC radial
profile is relatively insensitive to the value of $\beta$ for the
extended emission, but prefers a steeper distribution than $\beta =
2/3$ which was assumed by \markcite{worr94}Worrall \& Birkinshaw
(1994).  The radio structure is shown in \markcite{deruit86}de Ruiter
et al.~(1986) to be that of a weak, one-sided, jet within a diffuse,
low surface-brightness, envelope. The minimum pressure in different
parts of the radio source is compared with the gas pressure based on
the X-ray model in Figure~\ref{pressfig}.

\noindent{\bf 1040+31}. In \markcite{can99}Canosa et al.~(1999)
our spatial decomposition of the PSPC data uses a smaller on-source
region to probe the unresolved X-ray component.
There are no ROSAT HRI observations of this source.
\markcite{parm86}Parma et al.~(1986)
give a map of this source that shows it to
have a strong core, a small one-sided jet, fainter lobes, and several
``warm spots''. The minimum pressure in the extended emission,
estimated by Parma et al.~based on their map is (scaled to our
cosmology) about $2 \times 10^{-13}$~N~m$^{-2}$, or about
double the central pressure in the X-ray emitting gas. The radio
source has a total angular size of about 40~arcsec, roughly equal to
the core radius of the X-ray emission.

\noindent{\bf 1855+37}. In \markcite{can99}Canosa et al.~(1999) our
spatial decomposition of the PSPC data uses a smaller on-source region
to probe the unresolved X-ray component, a particularly small fraction
of the overall emission.  There are no ROSAT HRI observations of this
source.  The radio structure has been described by
\markcite{parm86}Parma et al.~(1986).  It appears as a small
(10~arcsec) double source, with a weak additional component to the
south. The minimum pressure in the source is approximately $3 \times
10^{-12}$~N~m$^{-2}$, comparable with the central pressure implied by
the X-ray model, but no information is available on structures less
than about 4~kpc in size, which may have higher pressures. Thus it
appears that, like 4C~35.03, the extended structure of this source is
limited by external gas pressure, but any jets which may be present on
smaller scales are likely to be overpressured.

\noindent{\bf 3C 449}. The spatial decomposition of the ROSAT HRI data
(\markcite{can99}Canosa et al.~1999) finds less extended emission than
that that reported here from the PSPC (understandable due to the lower
sensitivity of the HRI), while the unresolved emission of $10^{41}$
ergs s$^{-1}$ is in excellent agreement (see Tables~\ref{radialtab} \&
\ref{spectraltab}). The pressure balance between radio structures and
ROSAT PSPC-measured X-ray emitting gas is discussed by
\markcite{hard98}Hardcastle et al.~(1998), who in particular point out
the implications of an apparently underpressured southern radio lobe
displacing X-ray emitting plasma.  The minimum pressures in the radio
jets and lobes of 3C~449 were calculated and compared with the
pressure of the X-ray emitting atmosphere by
\markcite{hard98}Hardcastle et al.~(1998).  The conclusion of that
paper still stands: the entire source appears underpressured relative
to the external medium.  A jet angle $< 8^\circ$ relative to the line
of sight is needed if projection effects are to redress this balance,
but this seems unlikely.

\clearpage

\clearpage

\clearpage

\clearpage

\clearpage

\clearpage

\clearpage

\clearpage

\clearpage

\clearpage

\clearpage


\begin{deluxetable}{lllrcrc}
\tablewidth{0pt}
\tablecaption{Sample
\label{sampletab}}
\tablehead{
\colhead{B2 name}     & \colhead{Other name}  &
\colhead{$z$} & \colhead{$N_{\rm H}$ ($10^{20}$ cm$^{-2}$)$^{\rm a}$} 
& \colhead{kpc/arcmin$^{\rm b}$} & \colhead{LAS (arcsec)$^{\rm c}$} & \colhead{ref}
}
\startdata
0055+30&NGC 315&0.0165&5.77&28& 3650 & 1\nl
0055+26&NGC 326&0.0472&5.47&77&150 & 2\nl
0206+35&4C 35.03&0.0375&5.90&62&89&3\nl
0326+39&&0.0243&14.21&41&340&2\nl
0755+37&NGC 2484&0.0413&5.02&68&138&3\nl
1040+31& &0.036&1.82&60&54&3\nl
1855+37&&0.055&8.01&89&10&3\nl
2229+39&3C 449&0.0171&11.05&29&840&2\nl
\tablenotetext{a}{From \markcite{starke92}Stark et al. (1992).}
\tablenotetext{b}{$H_o$ = 50 km s$^{-1}$ Mpc$^{-1}$, $q_o$ = 0.}
\tablenotetext{c}{Largest angular size of radio structure}
\tablerefs{
(1) \markcite{bridle79}Bridle et al.~1979; 
(2) \markcite{ek81}Ekers et al.~1981; 
(3) \markcite{fant87}Fanti et al.~1987}
\enddata
\end{deluxetable}

\clearpage

\begin{deluxetable}{llcrcr}
\tablewidth{0pt}
\tablecaption{ROSAT PSPC Exposures
\label{exposuretab}}
\tablehead{
\colhead{Source}     & \colhead{Dates}  & \colhead{ROR$^{\rm a}$} &
\colhead{Exp.(s)$^{\rm b}$} & \colhead{Radius (arcmin)}
& \colhead{net counts$^{\rm c}$} 
}
\startdata
NGC 315&1992 Jan 17--Feb 1&700424&10,343&3&$500\pm28$\nl
$''$ &1992 Jul 19--21&$''$&17,869&3&$865\pm37$\nl
$''$ &total&$''$&29,626&3&$1430\pm63$\nl
NGC 326&1992 Jul 24--29&700884&19,409&6&$1066\pm42$\nl
4C 35.03&1992 Jul 24--27&700316&14,843&6&$512\pm41$\nl
0326+39&1993 Aug 15--25&701442&18,931&5&$933\pm65$\nl
NGC 2484&1991 Oct 30--Nov 1&700315&15,172&3&$523\pm36$\nl
1040+31&1993 May 10--21 &700883&21,952&6&$1492\pm68$\nl
1855+37&1993 Sep 27-28 &701445& 8,771&8&$3216\pm80$\nl
3C 449&1993 Jan 4--10 &700886&9,151&10&$1774\pm74$\nl
\tablenotetext{a}{Rosat Observation Request number.}
\tablenotetext{b}{Livetime used in analysis. May be shorter than
in the distributed files due to additional screening for
high background.}
\tablenotetext{c}{0.2-1.9 keV in source-centered circle of radius
given in column 5, after areas of any contaminating point sources
removed from on-source and background regions.  For NGC~326, counts at
radii $>$ 3 arcmin are for position angles 125$^\circ$--290$^\circ$
only, i.e., region of least extent.}
\enddata
\end{deluxetable}

\clearpage

\begin{deluxetable}{llllrlrc}
\tablewidth{0pt}
\tablecaption{Best-fit two-component models to the radial profiles
\label{radialtab}}
\tablehead{
\colhead{Source}     & \colhead{$C_{\rm u}/C_{\rm t}^{\rm \quad a}$}  
& \colhead{FWHM (kpc)$^{\rm ~b}$} &
\colhead{$\beta$} & \colhead{$\theta_{\rm cx}''$} 
& \colhead{$\chi^2/$dof}
& \colhead{$F(1,{\rm dof})^{\rm ~c}$}
& \colhead{$P_F(F,1,{\rm dof})^{\rm ~d}$}
}
\startdata
NGC 315$^{\rm e}$&$0.69\pm0.06$&$\phantom{4}20^{+8}_{-13}$&0.9&35&14.3/13&3.6&0.08\nl
$''$ &$0.46\pm0.05$&$\phantom{41}7^{+11}_{-4}$&0.67&10&13.9/13&2.5&0.14\nl
$''$ &--&$4.3\pm0.5$&0.67&6&14.6/14&--&--\nl
NGC 326&$0.06\pm0.01$&$417\pm40^{\rm~f}$&0.9&267&9.4/7&24.8&$0.002$\nl
4C 35.03&$0.18\pm0.03$&$99\pm83$&0.35&30&7.1/10&19.2&0.001\nl
0326+39&$0.14\pm0.02$&$130\pm50$&0.35&60&11.8/17&43.3&$<0.001$\nl
NGC 2484&$0.65\pm0.06$&$193^{+80}_{-110}$&0.9&140&12.4/12&8.7&0.01\nl
1040+31&$0.10\pm0.01$&$104^{+36}_{-69}$&0.4&40&14.3/22&10.9&0.003\nl
1855+37&$0.02\pm0.01$&$270\pm33$&0.4&70&19.3/20&22.1&$<0.001$\nl
3C 449&$0.02\pm0.01$&$\phantom{4}61\pm23$&0.35&40&7.5/12&8.9&0.01\nl
\tablenotetext{a}{Ratio of counts in nuclear unresolved component to the
total. $1\sigma$ statistical uncertainties.}
\tablenotetext{b}{FWHM of $\beta$-model component
($2\theta_{\rm cx}\sqrt{(2^{2\over 6 \beta -1}-1)}$).
90\% uncertainties for one interesting parameter.}
\tablenotetext{c}{F statistic, which tests the improvement of adding
the unresolved emission to a $\beta$-model alone.}
\tablenotetext{d}{Random probability of exceeding F.}
\tablenotetext{e}{Results for January, July, and combined data listed
separately.}
\tablenotetext{f}{Extended emission very asymmetric -- results are for
direction of least extent (see \markcite{wbc95}Worrall et al.~[1995]
for details).}
\tablecomments{Best-fit values for $0.35 \leq \beta \leq
0.9$. $\beta$ and $\theta_{\rm cx}$ are highly correlated and so
uncertainties are not given but can be deduced from Figure~\ref{chifig}.}
\enddata
\end{deluxetable}

\clearpage

\begin{deluxetable}{lccccccccc}
\tablewidth{0pt}
\tablecaption{Raymond-Smith Thermal Spectral Parameters
\label{spectraltab}}
\footnotesize
\tablehead{
&\multicolumn{3}{c}{Thermal Model} &
\multicolumn{4}{c}{Thermal + Power-Law Model} 
&\multicolumn{2}{c}{Total Thermal} 
\\
\cline{2-4} \cline{5-8} \cline{9-10} \\
&&&&&&&&\multicolumn{2}{c}{0.2-2.5 keV}\\
\cline{9-10} \\
&\colhead{$kT$} &  \colhead{norm$^{\rm a}$} &  \colhead{$\chi^2$}& 
\colhead{$f^{\rm b}$} & \colhead{$kT$} &  \colhead{norm$^{\rm a}$} &
\colhead{$\chi^2$} & \colhead{flux} & \colhead{$L$} \\
&\colhead{(keV)} &  \colhead{EM/$4\pi D_{\rm L}^2$}
&\colhead{/dof}
&&\colhead{(keV)}&\colhead{EM/$4\pi D_{\rm
L}^2$}&\colhead{/dof}&\colhead{$10^{-13}$}&\colhead{$10^{42}$}\\
source&& \colhead{$10^{10}$ cm$^{-5}$}
&&&&\colhead{$10^{10}$ cm$^{-5}$}&&\colhead{(cgs)}
&\colhead{(ergs/s)}\\
}
\startdata
NGC 315&$0.91^{+0.04}_{-0.04}$&$7.5^{+0.6}_{-0.6}$&49.9/30&
$0.39^{+0.16}_{-0.09}$&$0.62^{+0.09}_{-0.1}$&$3.44^{+0.59}_{-0.56}$&20.2/28&3.0&
$0.54\pm 0.2$\nl
NGC 326&$1.9^{+0.9}_{-0.4}$&$40.0^{+1.6}_{-2.3}$&25.3/29&
--&--&--&--&24&
$32.3\pm 1.3$\nl
4C 35.03&$1.4^{+1.3}_{-0.3}$&$7.6^{+0.9}_{-1.1}$&27.2/30&
$0.6^{+0.39}_{-0.6}$&$1.02^{+3.16}_{-0.59}$&
$2.43^{+5.8}_{-2.35}$&26.2/28&5.1&
$4.5\pm 0.4$\nl
0326+39&$1.02^{+0.12}_{-0.1}$&$11.6^{+1.3}_{-1.4}$&23.0/22&
$0.48^{+0.34}_{-0.32}$&$0.84^{+0.36}_{-0.32}$&$4.8^{+6.7}_{-3.1}$&20.2/20&8.5&
$4.1\pm 0.3$\nl
NGC 2484&$1.02^{+0.15}_{-0.12}$&$5.3^{+0.9}_{-0.85}$&30.0/29&
$0.64^{+0.29}_{-0.25}$&$0.73^{+0.32}_{-0.45}$&$1.46^{+2.34}_{-0.78}$&21.1/27&1.8&
$1.9\pm 0.2$\nl
1040+31&$1.21^{+0.16}_{-0.11}$&$10.9^{+0.8}_{-0.9}$&29.6/30&--&--&--&--&10&
$6.8\pm 0.3$\nl
1855+37&$2.16^{+0.7}_{-0.4}$&$91.7^{+3.1}_{-3.4}$&25.2/26&
--&--&--&--&67&
$152\pm 4$\nl
3C 449&$1.14^{+0.09}_{-0.08}$&$46.9^{+3.6}_{-3.6}$&21.7/30&
--&--&--&--&34&
$7.3\pm 0.3$\nl
\tablenotetext{a}{EM is volume-weighted emission measure ($\int n_e
n_p dV$) and $D_{\rm L}$ is luminosity distance.}
\tablenotetext{b}{Best-fit fraction of 0.2 - 1.9 keV counts in the
power law (error $\sim 90$\% confidence).}
\tablecomments{Fits assume Galactic $H_{\rm H}$ and abundances of 0.3
solar.  Errors in $kT$ and EM/$4\pi D_{\rm L}^2$ correspond to $\chi^2
+ 2.3$ (i.e. $1\sigma$ for 2 interesting parameters). For NGC~326,
results exclude the central excess and are taken from
\markcite{wbc95}Worrall et al.~(1995). For sources
where spatial fitting gives $\geq 14\%$ of the counts in a central
component, we investigate how the thermal parameters are affected if a
power-law component is included.  Flux and luminosity are best
overall estimates for the total thermal component.  The $1\sigma$ error on the
luminosity combines in quadrature the statistical error and that
arising from the uncertainty in the fraction of unresolved emission,
but does not take into account an uncertainty in the correction for
flux beyond the on-source extraction region, made using the best-fit
core radius and $\beta$ (or $\beta = 2/3$ and corresponding best-fit
core radius if $\beta_{\rm best}
\leq 0.5$).}  \enddata
\end{deluxetable}

\clearpage

\begin{deluxetable}{lclclccc}
\tablewidth{0pt}
\tablecaption{Cluster Central Gas Density, Pressure, and Cooling Time
\label{coolingtab}}
\tablehead{
\colhead{Source}     & \colhead{$B_o$}  & \colhead{$\beta$} &
\colhead{$r_{\rm cx}$} & \colhead{$kT$} &
\colhead{$n_{p,o}$} & \colhead{$P_o$}
& \colhead{$\tau_{\rm cool,~o}$}\\
& \colhead{cts arcmin$^{-2}$ ks$^{-1}$}  & &
\colhead{kpc} & \colhead{(keV)} &
\colhead{cm$^{-3}$} & \colhead{dynes cm$^{-2}$}
& \colhead{years}
}
\startdata
NGC 315&172$\pm$11&0.67&4.7&$0.62^{+0.09}_{-0.1}$&$3.5 \times 10^{-2}$&
$7.7 \times 10^{-11}$&$6.8 \times 10^{8}$\nl
NGC 326&\phantom{1}2.0$\pm$0.05&0.9&343&$1.9^{+0.9}_{-0.4}$&$7.3 \times 10^{-4}$&
$5.0 \times 10^{-12}$&$5.7 \times 10^{10}$\nl
4C 35.03&\phantom{1}3.7$\pm$0.3&0.35&\phantom{1}31&$1.3^{+1.4}_{-0.5}$&$2.0 \times 10^{-3}$&
$9.6 \times 10^{-12}$&$1.7 \times 10^{10}$\nl
0326+39&\phantom{1}3.9$\pm$0.2&0.35&\phantom{1}41&$0.95^{+0.36}_{-0.32}$&$1.8 \times 10^{-3}$&
$6.6 \times 10^{-12}$&$1.7 \times 10^{10}$\nl
NGC 2484&\phantom{1}1.5$\pm$0.2&0.9&159&$0.73^{+0.32}_{-0.45}$&$6.4 \times 10^{-4}$&
$1.7 \times 10^{-12}$&$4.0 \times 10^{10}$\nl
1040+31&\phantom{1}7.0$\pm$0.3&0.4&\phantom{1}40&$1.21^{+0.16}_{-0.11}$&
$2.3 \times 10^{-3}$&$1.0 \times 10^{-11}$&$1.4 \times 10^{10}$\nl
1855+37&17.5$\pm$0.4&0.4&104&$2.16^{+0.7}_{-0.4}$&$2.9 \times 10^{-3}$&
$2.3 \times 10^{-11}$&$1.5 \times 10^{10}$\nl
3C 449&10.0$\pm$0.3&0.35&\phantom{1}19&$1.14^{+0.09}_{-0.08}$&$4.4 \times 10^{-3}$&
$1.8 \times 10^{-11}$&$7.4 \times 10^{9}$\nl
\tablecomments{Based on $\beta$-model component only. 
The model normalization, $B_o$, is converted to
physical quantities by convolving a thermal spectrum
(with temperature given in Column~5 and based on
Table~\ref{spectraltab}) with the 
instrument response.
Density and pressure decrease with radius from
the tabulated central values as $\left(1 + {r^2 \over r^2_{cx}}
\right)^{-3\beta/2}$, and cooling time is proportional to 1/density.
Multiply pressure values by 0.1 to give in units of 
N m$^{-2}$ (Pascals).  The result for NGC~315 is the best based on
both epochs of PSPC data.}  

\enddata
\end{deluxetable}

\end{document}